\documentclass[runningheads]{svmult}

\usepackage{makeidx}   
\usepackage{graphicx}  
\usepackage{subeqnar}  
\usepackage{multicol}  
\usepackage{physprbb}  
\makeindex             

\begin{document}
\title*{Clustering at high redshift}
\toctitle{Clustering at high redshift}
%
%
\titlerunning{Clustering at high redshift}
%
\author{S. Cristiani\inst{1,2}
\and S. Arnouts\inst{3}
\and A. Fontana\inst{4}
\and P. Saracco\inst{5}
\and E. Vanzella\inst{3,6}
}
\authorrunning{Stefano Cristiani et al.}
%
%
\institute{ ST European Coordinating Facility, European Southern Observatory,
K.-Schwarzschild-Strasse 2, D-85748 Garching bei Muenchen, Germany  
\and Osservatorio Astronomico di Trieste, via Tiepolo 11, I-34131
Trieste, Italy 
\and European Southern Observatory, K.-Schwarzschild-Strasse 2,
D-85748 Garching bei Muenchen, Germany
\and Osservatorio Astronomico di Roma, via
dell'Osservatorio 2, Monteporzio, Italy 
\and Osservatorio Astronomico di Brera, via E. Bianchi 46,
Merate, Italy
\and Dipartimento di Astronomia, Universit\`a di Padova, vicolo
  dell'Osservatorio 2, I-35122 Padova, Italy
}

\maketitle              
\section{Near IR imaging of the HDF-S with VLT/ISAAC}
The addition of deep near infrared images to the database provided by
the HDF WFPC2 is essential to monitor the SEDs of the objects on
a wide baseline and address a number of key issues including the total
stellar content of baryonic mass, the effects of dust extinction, the
dependence of morphology on the rest frame wavelength, the photometric
redshifts, the detection and nature of extremely red objects (EROs).
For these reasons deep near infrared images were obtained with the
ISAAC instrument at the ESO VLT in the Js, H and Ks bands reaching,
respectively, 23.5, 22.0, 22.0 limiting Vega-magnitude 
($5 \sigma$ in an aperture of diameter $1^{``}.2 \equiv 2 \times \rm
{FWHM}$, \cite{saracco01,vanzella01}).
\section{A multi-color catalog of the HDF-S}
A multi-color (F300, F450, F606, F814, Js, H, Ks) photometric catalog 
of the HDF-S has been produced \cite{vanzella01} developing specific
procedures to match HST and VLT data. Having in mind the generation of
photometric redshifts we have chosen a conservative
approach in the object detection, leading to a list of
1611 sources. After correcting for the incompleteness of the source
counts, the object
surface density at $I_{AB} \le 27.5$ is estimated to be $220 ~ {\rm
arcmin}^{-2}$.  
The comparison between the median V-I colour in the HDF-North
and South shows a significant difference around $I_{AB} \sim 26$,
possibly due to the presence of large scale structure at $z \sim 1$ 
in the HDF-N.

Using for the object detection the Ks-band image we have  
selected down to $K_{AB} < 24$ a sample of 15 EROs,
defined as sources with $(I-K)_{AB} > 2.7$, corresponding  to the colour of
passively evolving elliptical galaxies at $z > 1$.
The EROs surface density turns out to be $3.2 \pm 0.9  ~ {\rm
arcmin}^{-2}$, 
and their distribution, at least from the angular point 
of view, is remarkably nonuniform:
10 EROs out of 15 are inside the upper WFPC2 chip (0.6 square arc-minute). 
One of the EROs is a powerful radio-galaxy.
\section{Photometric and Spectroscopic Redshifts}
Photometric redshifts have been produced both fitting templates to the 
observed SEDs \cite{arnouts01,fontana02} and with neural network
techniques \cite{vanzella02}.
Using colour-colour diagrams, 90 U-band dropouts have been
selected down to $I_{AB} =27$ (19 are brighter than 25 mag).
Spectroscopic observations of the 9 candidates with $I_{AB} <24.5$ have
been carried out, confirming all of them to be galaxies with $2<z<3.5$.
Similarly, 17 B-band dropouts have been selected down to
$I_{AB}=27$ (all with $I_{AB}>26$).
A comparison of the 38 spectroscopic redshifts available so far with
the photometric predictions, provides an estimate of the redshift accuracy
$\sigma_z$ of 0.16, 0.13 for the template fitting and neural network
technique, respectively.
\section{The Redshift Evolution of the Galaxy Clustering} 
The photometric redshifts for all the galaxies brighter than $I_{AB}<
27.5$ have been used to study the evolution of galaxy clustering in
the interval $0<z<4.5$ \cite{arnouts01}. 
The clustering signal is obtained in different redshift bins using two
different approaches: a standard one, which uses the best redshift
estimate of each object, and a second one, which takes into account
the redshift probability function of each object and 
improves the information in the
redshift intervals where the contamination from objects with insecure
redshifts is important.  
With both methods, we find that the clustering strength up to $z \sim 3.5$ in
the HDF-S is consistent with the previous results in the HDF-N. 
While at redshift lower than $z \sim 1$ the HDF galaxy population is 
un/anti-biased ($b<1$) with respect to the underlying dark matter, 
at high redshift the bias increases up to $b\sim 2-3$, depending on 
the cosmological model. 
These results support previous claims that, at high redshift, galaxies
are preferentially located in massive haloes, 
as predicted by the biased galaxy formation scenario. 
The impact of cosmic errors on the analysis has been quantified, 
showing that errors in the clustering measurements in the HDF surveys
are indeed dominated by shot-noise in most regimes. 
Future observations with instruments like the ACS on HST will improve
the S/N by at least a factor of two and more detailed analyses of the 
errors will be required. In fact, pure shot-noise will give a smaller
contribution with respect to other sources of errors, such as finite 
volume effects or non-Poissonian discreteness effects.

%

\end{document}